\documentclass[aps,draft,preprint,groupedaddress]{revtex4}
\newcommand{\beq}{\begin{equation}}
\newcommand{\eeq}{\end{equation}}
\newcommand{\beqn}{\begin{eqnarray}}
\newcommand{\eeqn}{\end{eqnarray}}

\newcommand{\al}{Z\alpha}
\newcommand{\alv}{\mbox{\boldmath $\alpha$ \unboldmath}}
\newcommand{\eps}{\varepsilon}
\newcommand{\g}{\mbox{\boldmath ${\gamma}$\unboldmath}}

\newcommand{\p}{\mbox{${\bf p}$}}

\newcommand{\ro}{\mbox{\boldmath ${\rho}$\unboldmath}}

\newcommand{\q}{{\bf q}}
\newcommand{\vd}{\mbox{${\bf \Delta}$}}

\begin{document}
\bibliographystyle{apsrev}
%\preprint{BINP 2001-?}

\title{Coulomb corrections and multiple $e^+e^-$ pair production
in ultra-relativistic nuclear collisions}
\author{R.N. Lee}
\email[Email:]{R.N.Lee@inp.nsk.su}
\author{A.I. Milstein}
\email[Email:]{A.I.Milstein@inp.nsk.su} \affiliation{Budker
Institute of Nuclear Physics, 630090 Novosibirsk, Russia}

\date{March 20, 2001}
\begin{abstract}
We consider the problem of Coulomb corrections to the inclusive
cross section. We show that these corrections in the limiting
case of small charge number of one of the nuclei coincide with
those to the exclusive cross section. Within our approach we
also obtain the Coulomb corrections for the case of large charge
numbers of both nuclei.
\end{abstract}
\pacs{} \maketitle

In a set of recent publications the process of $e^+e^-$ pair
production in ultrarelativistic heavy-ion collisions was
investigated by different groups of authors. The authors of
\cite{SW,McL,Gre} treated the nuclei as sources of the external
field, and calculated the amplitude of the process at a fixed
impact parameter using retarded solutions of the Dirac equation.
After that the cross section was obtained by the integration
over the impact parameter:
\begin{eqnarray}
\label{section1} d\sigma&=&
\frac{m^2d^3p\,d^3q}{(2\pi)^6\eps_p\eps_q} \int d^2\rho
\left|\int \frac{d^2k}{(2\pi)^2} \exp[i\mathbf{k}\ro]{\cal M}\,
{\cal F}_A(\mathbf{k}')\,{\cal
F}_B(\mathbf{k})\right|^2 \\
{\cal M}&=&\overline{u}(p)\left[\frac{\alv (\mathbf{k}-\p_\perp)
+ \gamma_0 m}{-p_+ q_- - (\mathbf{k}-\p_\perp)^2 -m^2 } \gamma_-
+\frac{-\alv (\mathbf{k}-{\q}_\perp) + \gamma_0 m}{-p_- q_+ -
(\mathbf{k}-{\q}_\perp)^2 -m^2} \gamma_+\right]u(-q)\,
.\nonumber
\end{eqnarray}
Here $\p$ and $\eps_p$ ($\q$ and $\eps_q$) are the momentum and
energy of the electron (positron), $u(p)$ and $u(-q)$ are
positive- and negative-energy Dirac spinors, $\alv=\gamma^0\g$,
$\gamma_\pm=\gamma^0\pm\gamma^z$, $\gamma^\mu$ are the Dirac
matrices, $p_\pm=\eps_p\pm p^z$, $q_\pm=\eps_q\pm q^z$, $m$ is
the electron mass, $\mathbf{k}$ is a two-dimensional vector
lying in the $xy$ plane,
$\mathbf{k}'=\q_\perp+\p_\perp-\mathbf{k}$, and the function
${\cal F}(\vd)$ is proportional to the electron eikonal
scattering amplitude in the Coulomb field:
\beq\label{impact1}
{\cal F}(\vd)= i\pi\al \frac{\Gamma(1-i\al)}{\Gamma(1+i\al)}
\left(\frac{4}{\Delta^2}\right)^{1-iZ\alpha} \ ,
\eeq
where $Z=Z_{A,B}$ is the charge number of the nucleus $A,B$. The
nuclei $A$ and $B$ are assumed to move in the positive and
negative directions of the $z$ axis, respectively, and have the
Lorentz factors $\gamma_{A,B}=1/\sqrt{1-\beta_{A,B}^2}$. Using
(\ref{section1}) the authors of \cite{SW,McL,Gre} made the
conclusion that the exact cross section coincides with that
calculated in the Born approximation, i.e. in the lowest-order
perturbation theory with respect to $\alpha Z_{A,B}$. The Born
cross section can be obtained from (\ref{section1}) by the
replacement ${\cal F}(\vd)\to {\cal F}^0(\vd)=4i\pi
Z\alpha/\Delta^2$. The statement about the absence of the
Coulomb corrections was criticized in \cite{Ser}, where in the
frame of the Weizs\"acker-Williams approximation with respect to
one of the nucleus the cross section of the process was
expressed via the cross section of $e^+e^-$ pair production by a
photon in a Coulomb field \cite{Bet}. As is well known, the
latter contains the Coulomb corrections (higher order terms in
$Z\alpha$). In our paper \cite{LeeM1} we explicitly demonstrated
that the following statements are true for the results obtained
in \cite{SW,McL,Gre}:
\begin{itemize}
\item the expression (\ref{section1}) actually contains the Coulomb corrections. The wrong
conclusion on the absence of Coulomb corrections made
in \cite{SW,McL,Gre} is connected with illegal change of the
order of integration in repeated integrals before the
regularization of integrand.
\item it cannot be applied for the calculation of the differential cross section
with respect to both $e^+$ and $e^-$. The impossibility to
interpret (\ref{section1}) as the cross section differential
with respect to both particles is connected with the use of wave
functions having the improper asymptotic behavior for the
problem of pair production (retarded wave functions). This point
was later realized by the authors of \cite{EichmRG1}.
\item the cross section (\ref{section1}) calculated in the lowest order
in $Z_A\alpha$ ( proportional to $(Z_A \alpha)^2$) and
integrated over the momenta of at least one particle of the pair
contains the Coulomb corrections in $Z_B \alpha$ which are in
agreement with those obtained in the Weizs\"acker-Williams
approximation.
\end{itemize}

Though our paper \cite{LeeM1} essentially clarified the
situation with the Coulomb corrections, recently the paper
\cite{McL1} appeared. The authors of \cite{McL1} still claim
that the expression does not contain the Coulomb corrections.
They also pointed out that the expression (\ref{section1}) makes
sense only after the integration over the momenta $\mathbf{p}$
and $\mathbf{q}$ and gives the inclusive cross section
$\sigma_T$ of pair production being defined as
\begin{equation}
\sigma_T\equiv \int d^2{\ro}\; \sum_{n=1}^{+\infty}n P_n\, ,
\end{equation}
where $P_n$ is the probability to produce exactly $n$ pairs in a
collision at impact parameter ${\ro}$. Note that the usual
definition of the inclusive cross section, as a sum of cross
sections of all possible processes, is different:
\begin{equation}
\label{sigma_incl} \sigma_{incl}\equiv \int d^2{\ro}\;
\sum_{n=1}^{+\infty} P_n\, .
\end{equation}
The probabilities $P_n$ are calculated exactly in the external
field and their sum in the r.h.s. of (\ref{sigma_incl}) is
expressed via the vacuum-to-vacuum transition probability $P_0$
as $\sum_{n=1}^{\infty} P_n=1-P_0$.

The cross section $\sigma_T$ differs from the exclusive cross
section $\sigma_1$ of the production of exactly one pair
\begin{equation}
\sigma_1\equiv \int d^2{\ro}\; P_1\,
\end{equation}
The authors of \cite{McL1} suppose that this circumstance
justifies the absence of the Coulomb corrections in their
result.

In the present paper we consider the problem of Coulomb
corrections to $\sigma_T$ and demonstrate their existence.

Let us show first that difference in the definition of
$\sigma_T$ and $\sigma_1$ can not justify the absence of the
Coulomb corrections in the former. Indeed, the expansion of the
probability $P_n$ in the parameters $Z_A\alpha$ and $Z_B\alpha$
starts from the term, proportional to $(Z_A\alpha)^{2n}
(Z_B\alpha)^{2n}$. Therefore the terms $\propto (Z_A\alpha)^2
(Z_B\alpha)^{2l}$ with $l\geq 1$ are contained only in $P_1$.
Therefore, the terms, quadratic in $Z_A\alpha$, in $\sigma_T$
and $\sigma_1$ should coincide. By means of the
Weizs\"acker-Williams approximation with respect to the nucleus
$A$ it is easy to understand that the term in $\sigma_1$,
quadratic in $Z_A\alpha$, contains the Coulomb corrections in
the parameter $Z_B\alpha$.

Now we pass to the explicit calculations of $\sigma_T$. For the
sake of simplicity we consider the process in the frame where
both nuclei have the same Lorentz factor
$\gamma_A=\gamma_B=\gamma=1/\sqrt{1-\beta^2}$. First of all it
is worth noting that the integration of (\ref{section1}) with
respect to $\mathbf{p,q}$ and $\ro$ leads to the logarithmic
divergence. Of course, this is the consequence of setting
$\beta=1$ in the light-front approach used in \cite{SW,McL,Gre}.
In more accurate approach the regularizing terms proportional to
$1/\gamma^2$ should be kept in denominators. In particular, this
leads to the regularization of the integrals over $p_z$ and
$q_z$, which is equivalent (with the logarithmic accuracy) to
the integration over these variables from $-m\gamma$ to
$m\gamma$. For this momenta the velocities of both particles in
pair are less than those of the nuclei.

Since the expression (\ref{section1}) is ill-defined, it is
possible to perform the mathematical transformations of it only
after the regularization. Assuming this regularization to be
made in ${\cal M}$ and ${\cal F}$ we take the integral over
$\ro$. As a result we have for $\sigma_T$:

\begin{eqnarray}
\label{section2} \sigma_T&=& \int\!\!\int
\frac{m^2d^3p\,d^3q}{(2\pi)^6\eps_p\eps_q}\frac{d^2k}{(2\pi)^2}
|{\cal M}|^2 |{\cal F}_A(\mathbf{k}')|^2|{\cal
F}_B(\mathbf{k})|^2
\end{eqnarray}
If one substitutes non-regularized ${\cal F}$ from
(\ref{impact1}) into (\ref{section2}) then the Coulomb
corrections cancel. However, this substitution is illegal since
it leads to the divergence. It was the source of mistake made in
\cite{SW,McL,Gre,McL1}.

To proceed with the calculations it is convenient to split the
expression (\ref{section2}) as
\begin{equation}\label{splitted}
  \sigma_T=\sigma^b+\sigma^c_A+\sigma^c_B+\sigma^c_{AB}\, ,
\end{equation}
where
\begin{eqnarray}
\label{sectionsplit} \left.
\begin{array}{l}
  \sigma^b \\
  \sigma^c_{B} \\
  \sigma^c_{AB}
\end{array}
\right\} &=& \int\!\!\int
\frac{m^2d^3p\,d^3q}{(2\pi)^6\eps_p\eps_q}\frac{d^2k}{(2\pi)^2}
|{\cal M}|^2 \times\left\{\begin{array}{l}
   |{\cal F}^0_A(\mathbf{k}')|^2|{\cal
F}^0_B(\mathbf{k})|^2 \\
  |{\cal F}^0_A(\mathbf{k}')|^2\left[|{\cal
F}_B(\mathbf{k})|^2-|{\cal F}^0_B(\mathbf{k})|^2\right] \\
  \left[|{\cal F}_A(\mathbf{k}')|^2-|{\cal
F}^0_A(\mathbf{k}')|^2\right] \left[|{\cal
F}_B(\mathbf{k})|^2-|{\cal F}^0_B(\mathbf{k})|^2\right]
\end{array}\right.
\end{eqnarray}
The term $\sigma^c_A$ is obtained from $\sigma^c_B$ by obvious
substitution. Here ${\cal F}^0(\vd)$, as well as ${\cal
F}(\vd)$, is assumed to be regularized in a proper way. In
(\ref{splitted}) the term $\sigma^b$ is the Born part of
$\sigma_T$, $\sigma^c_{A}$ and $\sigma^c_{B}$ contain the terms
proportional to $(Z_B\alpha)^2 (Z_A\alpha)^{2n}$ and
$(Z_A\alpha)^2 (Z_B\alpha)^{2n}$, respectively, $n\geq 2\,$. At
last, $\sigma^c_{AB}$ contains the terms proportional to
$(Z_A\alpha)^{n} (Z_B\alpha)^{l}$ with $n,l>2\,$.

Let us discuss now the regularization. Note that the expression
(\ref{section1}) was derived without using specific character of
the Coulomb potential. For arbitrary potential $V(r)$ we have
\beqn\label{impact}
{\cal F}(\vd)&=&\int d^2\rho \exp[-i \ro \vd]
\left\{\exp[-i\chi(\rho)]-1\right\} \ ,\\
\chi(\rho)&=&\int\limits_{-\infty}^\infty dz
V\left(\sqrt{z^2+\rho^2}\right)\ . \nonumber
\eeqn
The eikonal phase $\chi(\rho)$ is finite if $r V(r)\to 0$ at
$r\to\infty$. Moreover, under this restriction on the potential
we obtain the finite result for $\sigma_T$. As known, the
correct expression for ${\cal F}^0_{A,B}(\vd)$ for $\beta<1$ is
\begin{eqnarray}\label{BornPropagators}
{\cal F}^0_{A,B}(\mathbf \Delta) &=& \frac{4i\pi
Z_{A,B}\alpha}{\Delta^2+a_\pm^2}\, ,\quad
a_\pm=(p_\pm+q_\pm)/2\gamma\, .
\end{eqnarray}
These expressions for ${\cal F}^0_{A,B}(\vd)$ correspond to the
choice of effective potential
$V_{A,B}(r)=-{Z_{A,B}\alpha}\exp\left[-r a_\pm \right]/r$. The
quantity ${\cal F}$ for this potential reads
\beqn\label{impact2}
{\cal F}_{A,B}(\vd)&=&2\pi\int d\rho\rho J_0(\rho \Delta)
\left\{\exp[2iZ_{A,B}\alpha K_0(\rho a_\pm)]-1\right\} \, ,
\eeqn
where $J_0$ is the Bessel function and $K_0$ is the modified
Bessel function of the third kind. Let us emphasize that the
regularization of $\cal F$ in (\ref{impact2}) is not reduced to
the substitution $\vd^2\to\vd^2+a_\pm^2$ in (\ref{impact1})
which was suggested in \cite{McL}.

For all terms in $\sigma_T$ the main contribution to the
integrals comes from the region of integration
$$
|\mathbf{k}|,\,|\mathbf{k}'|\ll m\,,\qquad|p_z|,\,|q_z|\ll
m\gamma\,,\qquad |\p_\perp-\q_\perp|\sim m\,.
$$
According to the first restriction we can expand ${\cal M}$ with
respect to both $\mathbf{k}$ and $\mathbf{k}'$. Due to the gauge
invariance the first non-zero term of this expansion reads
${\cal M}= k_ik'_jM_{ij}$. Passing to the variables
$\mathbf{k},\,\mathbf{k}'$, and
$\mathbf{r}=(\p_\perp-\q_\perp)/2$, we obtain from
(\ref{sectionsplit}):
\begin{eqnarray}
\label{sectionsplit1} \left.
\begin{array}{l}
  \sigma^b \\
  \sigma^c_A\\
  \sigma^c_B\\
  \sigma^c_{AB}
\end{array}\right\}&=& \int\!\!\int
\frac{m^2dp_z\,dq_zd^2r}{4(2\pi)^4\eps_p\eps_q} |M_{ij}|^2
\left\{
\begin{array}{l}
L_A L_B\\
G_A L_B\\
L_A G_B\\
G_A G_B
\end{array}\right.
\end{eqnarray}
Here
\begin{eqnarray}\label{LG}
 L_{A,B}&=&\int\limits_{|\mathbf{k}|<m}
 \frac{d^2k}{(2\pi)^2} k^2|{\cal F}^0_{A,B}(\mathbf{k})|^2=8\pi
 (Z_{A,B}\alpha)^2 \ln (m/a_\pm)\, ,\\
 G_{A,B}&=&\int
 \frac{d^2k}{(2\pi)^2}k^2 \left[|{\cal F}_{A,B}(\mathbf{k})|^2-|{\cal
F}^0_{A,B}(\mathbf{k})|^2\right] \, .\nonumber
\end{eqnarray}
The functions $G_{A,B}$ have been calculated in our paper
\cite{LeeM1}. It was shown in \cite{LeeM1} that, though the main
contribution to the integral over $\mathbf{k}$ in $G_{A,B}$
comes from the region $k\sim a_{\pm}\ll m$, where $|{\cal
F}^0_{A,B}(\mathbf{k})|$ differs from $|{\cal
F}_{A,B}(\mathbf{k})|$, the quantities $G_A$ and $G_B$ are
universal functions of $Z_A\alpha$ and $Z_B\alpha$,
respectively. They have the form
\beqn
\label{Gfin}
G_{A,B}&=&-8\pi(Z_{A,B}\alpha)^2[\mbox{Re}\,\psi(1+i
Z_{A,B}\alpha)+C] =-8\pi(Z_{A,B}\alpha)^2 f(Z_{A,B}\alpha)  \ ,
\eeqn
where $C$ is the Euler constant, $\psi(x)=d\ln\Gamma(x)/dx$.

Straightforward calculation leads to the following expression
for $|M_{ij}|^2$:
\begin{equation}\label{Mij2}
  |M_{ij}|^2=\frac{8}{m^2(r^2+m^2)(p_-+q_-)(p_++q_+)}\left[
1-\frac{2(r^4+m^4)}{(r^2+m^2)(p_-+q_-)(p_++q_+)}
  \right]\, ,
\end{equation}
where $p_\pm=\sqrt{p_z^2+r^2+m^2}\pm p_z$ and similar to
$q_\pm$. At the derivation of (\ref{Mij2}) we have performed the
summation over the polarizations of both particles in pair.
Substituting (\ref{Mij2}) in (\ref{sectionsplit1}) and
performing the integration with the logarithmic accuracy, we
obtain for~$\sigma^b$~and~$\sigma^c_{A,B}$:
\begin{eqnarray}\label{sectionbcba}
\sigma^b&=&\frac{28(Z_A\alpha)^2(Z_B\alpha)^2}{27\pi
m^2}\ln^3(\gamma^2)\,,\\
\sigma^c_{A}&=&-\frac{28(Z_A\alpha)^2(Z_B\alpha)^2}{9\pi
m^2}f(Z_A\alpha)\ln^2(\gamma^2)\,,\nonumber\\
\sigma^c_{B}&=&-\frac{28(Z_A\alpha)^2(Z_B\alpha)^2}{9\pi
m^2}f(Z_B\alpha)\ln^2(\gamma^2)\,.\nonumber
\end{eqnarray}
As expected, these results agree with those obtained with the
use of the Weizs\"acker-Williams approximation. Within our
approach we also get the following expression for
$\sigma^c_{AB}$:
\begin{equation}\label{sectionab}
  \sigma^c_{AB}=\frac{56(Z_A\alpha)^2(Z_B\alpha)^2}{9\pi
m^2}f(Z_A\alpha)f(Z_B\alpha)\ln(\gamma^2)
\end{equation}
In the arbitrary frame one should replace $\gamma^2\to
\gamma_A\gamma_B$ in (\ref{sectionbcba}) and (\ref{sectionab}).

Thus we demonstrated that the difference in definitions of the
exclusive cross section $\sigma_1$ and the inclusive cross
section $\sigma_T$ can not lead to the cancellation of the
Coulomb corrections in the latter. Using the proper
regularization of the expression for $\sigma_T$ we obtained the
result for Coulomb corrections which in the limiting case
$Z_A\alpha\ll 1$ (or $Z_B\alpha\ll 1$) agrees with that obtained
in the Weizs\"acker-Williams approximation.

\end{document}